\documentclass[twocolumn,showpacs,preprintnumbers, superscriptaddress,  longbibliography, amssymb, pra]{revtex4-1}

\usepackage{amsmath}
\usepackage{graphicx}
\usepackage{dcolumn}
\usepackage{bm}
\usepackage{verbatim}
\usepackage{array}
\usepackage{epsfig}
\usepackage{amsbsy}
\usepackage{color}
\usepackage{soul}
\usepackage{url}
\usepackage{hyperref}
\usepackage{longtable}

\usepackage[mediumqspace,mediumspace,amssymb,Gray]{SIunits}

\def\ket#1{\mathinner{|{#1}\rangle}}

\begin{document}
\preprint{\hfill\parbox[b]{0.3\hsize}{ }}

\title{State-selective influence of the Breit interaction on the \\  angular distribution of emitted photons following dielectronic recombination}



\author{Pedro Amaro} 
\altaffiliation[Now at ]
{LIBPhys-UNL,  FCT-UNL,  P-2829-516, Caparica, Portugal}
\email{pdamaro@fct.unl.pt}
\affiliation{Physikalisches Institut,~Universit\"at Heidelberg,~D-69120 Heidelberg,~Germany}

\author{Chintan Shah}
\affiliation{Physikalisches Institut,~Universit\"at Heidelberg,~D-69120 Heidelberg,~Germany}
\affiliation{Max-Planck-Institut f\"{u}r Kernphysik,~D-69117 Heidelberg, Germany}

\author{Rene Steinbr\"{u}gge}
\altaffiliation[Now at ]
{TRIUMF, 4004 Wesbrook Mall, Vancouver, Canada.}
\affiliation{Max-Planck-Institut f\"{u}r Kernphysik,~D-69117 Heidelberg, Germany}

\author{Christian Beilmann}
\altaffiliation[Now at ]
{Karlsruhe Institute of Technology, 76131 Karlsruhe,  \linebreak Germany.}
\affiliation{Physikalisches Institut,~Universit\"at Heidelberg,~D-69120 Heidelberg,~Germany}
\affiliation{Max-Planck-Institut f\"{u}r Kernphysik,~D-69117 Heidelberg, Germany}

\author{Sven Bernitt}
\altaffiliation[Now at ]
{ Institut f\"{u}r Optik und Quantenelektronik, D-07743 Jena,  \linebreak Germany.}
\affiliation{Max-Planck-Institut f\"{u}r Kernphysik,~D-69117 Heidelberg, Germany}



\author{Jos\'e R. Crespo L\'opez-Urrutia}
\affiliation{Max-Planck-Institut f\"{u}r Kernphysik,~D-69117 Heidelberg, Germany}

\author{Stanislav Tashenov}
\affiliation{Physikalisches Institut,~Universit\"at Heidelberg,~D-69120 Heidelberg,~Germany}

\date{Received: \today  }


\begin{abstract}

We report a measurement of $KLL$  dielectronic recombination in charge states from Kr$^{+34}$ through Kr$^{+28}$, in order to investigate the contribution of Breit interaction for a wide range of resonant states.
%
Highly charged Kr ions were produced in an electron
beam ion trap, while the electron-ion collision energy was scanned over a
range of dielectronic recombination resonances. 
The subsequent $K\alpha$ x rays were recorded both along and perpendicular to the
electron beam axis, which allowed the observation of the  influence of Breit interaction on the angular distribution of the x rays. Experimental results are in good agreement with distorted-wave calculations. 
We demonstrate, both theoretically and experimentally,  that there is a strong state-selective influence of the Breit interaction that can be traced back to the angular and radial properties of the wavefunctions in the dielectronic capture.




\end{abstract}


\maketitle


\section{Introduction}
\label{intro}

Highly charged ions are ideal atomic systems for investigating the relativistic details of dynamical atomic processes, such as electron-ion collisions.  At the strong electromagnetic fields of these atomic systems, both incident and bound electrons can reach sizable fractions of the light speed, and consequently, retardation and magnetic terms of the electron-electron interaction can significantly change the collision dynamics. 
These relativistic terms are included in the Breit interaction \cite{bre1930,bre1932}, which corresponds to the next quantum electrodynamic term after Coulomb interaction \cite{suc1980, sap1987}. The contribution of the Breit interaction to the electronic structure has been extensively calculated, using various methods \cite{bre1930,bre1932,sap1998, joh2007},   showing  excellent agreement with experimental results \cite{gtb2005, fis2005}. 
Its typical contribution to the binding energy of a few percent  \cite{mjo1971,grp1976}  can be treated perturbatively. 
 On the other hand, the electron-ion collision dynamics is much more dependent on the electron-electron interaction, which hampers perturbative approaches to the Breit contribution. 
Hence, recent experiments  have  been performed in several electron-ion atomic processes, namely, resonant transfer and excitation \cite{mmb2003}, Coulomb excitation \cite{gfh2011} and electron impact excitation \cite{gtf2013} in order to investigate the relativistic details of the respective atomic collisions. 
%

Dielectronic recombination (DR) provides an even more sensitive probe of  the Breit interaction. In this process, a free electron is captured by an ion with simultaneous excitation of a bound electron, which only occurs due to the electron-electron interaction.  Deexcitation of the resonant state by photon emission  completes the DR process.
Recent investigations highlighted a pronounced  role of the Breit interaction along the (usually dominant) Coulomb interaction, not only on the resonant strengths \cite{nkw2008}, but also on the angular distribution and polarization properties of the photon emission \cite{fss2009, sjb2015}. 
 All of these investigations focused on one particular resonant state, $[1s 2s^22p_{1/2}]_1$, of initial Li-like ion. Apart from this unique case, only few resonant DR states of initial H-like ions were  reported also having a dominant Breit influence \cite{csc1995, bbc2011}.

Besides this fundamental interest on the relativistic details of electron-electron interaction, investigations of the DR are also mandatory for modeling hot astrophysical and laboratory plasmas, as DR often dominates the recombination rates due to its high resonance strength \cite{olb2010, mhs2011, zcg2004}.  Resonances can thus be used for temperature and density plasma diagnostics \cite{bur1964, bhb1993, wbd1995, see1979}. Simultaneously, information about the angular distribution and polarization of the DR photons emitted by anisotropic plasmas could be used to infer the directionality of the plasma electrons. Moreover, an anisotropic emission may change the intensity of observed lines and can thus affect temperature diagnostics \cite{duv1980, smn1998, smn1999,  ind1987,  ind1989, diu1996}. Besides DR,  trielectronic and quadruelectric recombination was recently demonstrated  to play an essential role in  charge state distribution of hot plasmas \cite{sas2016}.

Recently, the polarization of DR $K\alpha$ x rays from highly charged Kr and Xe were measured with a novel  Compton polarimeter \cite{Jhb2015, sjb2015}. 
The degree of linear polarization was determined for a  few  DR resonances and showed agreement with theoretical predictions of a measurable influence of  the Breit interaction (BI). 
In this work, we complement those previous studies with a systematic investigation of the influence of BI on the angular distribution of emitted photons following DR for a wide range of DR resonances.  
For this purpose, as in earlier works, charge states of He-like ( Kr$^{+34}$) to O-like (Kr$^{+28}$) were produced in an electron beam ion trap  (EBIT), and the electron beam energy was scanned over the $KLL$ resonances. We measured the angular distribution of the subsequent $K\alpha$ x rays  through  simultaneous observation in directions  along and perpendicular to the electron beam axis.
 The alignment parameter of the resonant state was extracted from the acquired spectra. 
  These results were compared with extensive distorted-wave calculations based on the Flexible Atomic code (FAC).  
  Here, we focus only on the state-selective influence of the Breit interaction on the angular distribution, i.e., why some resonances display such strong BI influence, while others only manifest a negligible one. A previous study was  performed  for only  the two states $[1s2s^22p_{1/2}]_1$ and $[1s2s 2p_{1/2}^2]_1$ in Li-like ions \cite{thl2015}.
 %
 %

 \begin{figure}[t]
  \centering
\includegraphics[clip=true,width=1.0\columnwidth]{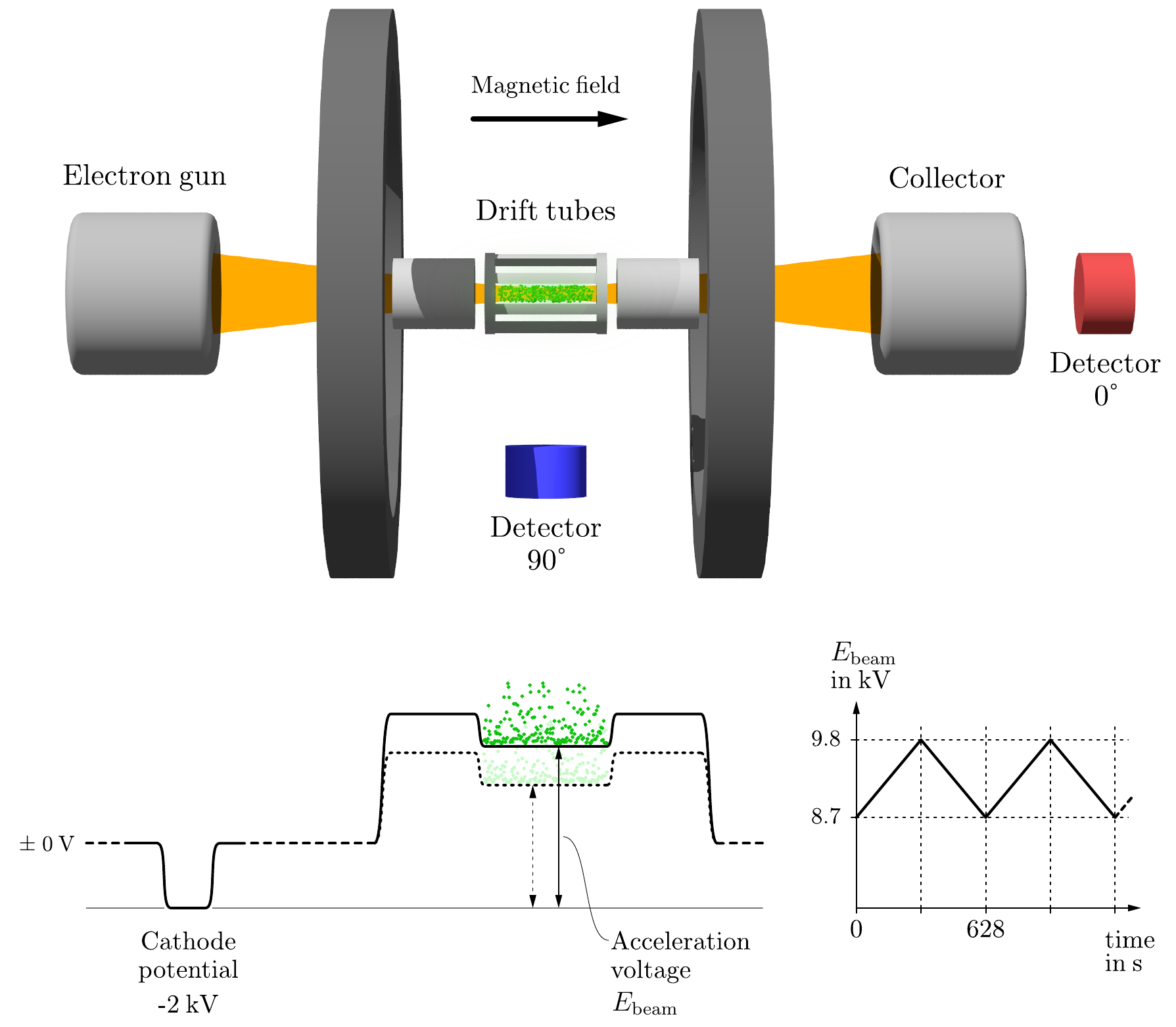}
\caption{ (Color online) Scheme of the experimental setup. An electron beam is accelerated toward the drift tubes and compressed by a magnetic field. Highly charged Kr ions are then produced in the trap by successive electron impact ionization. The electron beam energy is scanned over the DR resonances and the subsequent photon emission is observed by two x-ray detectors located along (0\degree) and perpendicular to the electron beam axis (90\degree).
} 
\label{fig:exp_ebit}
\end{figure}
%

\section{Theory}
\label{sec:theor}

We give here a brief description of the theoretical framework used for describing the DR in few-electron ions, as this  process has already been extensively  discussed  in the literature \cite{zhs2002, czh2015, fss2009, fkn2008}.  
%
%
  %
%
Due to the collision of unidirectional electrons with the ions in the present setup, the formed resonant states are aligned along the beam axis, i.e., the magnetic sublevel population of a resonant state $\ket{\alpha_d J_d M_d}$ is non-statistically distributed.  
Alignment parameters $\mathcal{A}_k ^d$ are often used to describe the photon angular distribution of non-statistically distributed states. For the DC process, $\mathcal{A}_k ^d$ has the following form:
\begin{eqnarray} 
&&\mathcal{A}_k ^d = \sqrt{2J_d+1} \nonumber \\
&&\times \sum_{M_d=-J_d}^{J_d} (-1)^{J_d-M_d} \left< J_d M_d, J_d-M_d | k~0\right>\bar{\sigma}_{M_d} ~.
\label{eq:alignm_cross}
\end{eqnarray} 
Here, $\bar{\sigma}_{M_d}$ are normalized magnetic sublevel  cross sections of the DC process ($\bar{\sigma}_{M_d}=\sigma_{M_d} /\sum_{M_d'}\sigma_{M_d'}$).  
The parameters $\mathcal{A}_k ^d$  have only nonzero values if $k$ is even and 
$k< J_d$. Therefore,  only resonant states with $J_d > 1/2$ can be aligned.
Aligned states often emits anisotropic radiation, and for the DR process, the angular distribution is given by \cite{bal2000}
%
 \begin{equation}
 \left(\frac{dS}{d\Omega} \right)_{d\rightarrow f} =  \frac{ S}{4 \pi} \left[1+ \sum_{k=2,4...} \alpha_{d\rightarrow f}^k \mathcal{A}_k^d P_k (\cos \theta)    \right] ~,
 \label{eq:ang_dis}
 \end{equation}
with $S$ being the DR resonance strength. The polar angle $\theta$ is defined with respect to the electron beam axis and $P_k$  denotes the Legendre polynomial. Information concerning the photon emission between the intermediate resonant  and final states, such as the multipole contributions, is included in  the intrinsic photon coefficients $\alpha_{d\rightarrow f}^k$ \cite{bal2000}. 
 All resonant  states listed in Table~\ref{tab:res_ratio} contains an allowed and dominant  electric dipole channel. Therefore, all terms $\alpha_{d\rightarrow f}^k$ with $k>2$ can be neglected  ($\alpha_{d\rightarrow f}^2 \equiv \alpha_{d\rightarrow f}$, see Ref.~\cite{sjb2015}).
 %
%
In this investigation, we do not consider a possible alignment of the initial state  that might arise from the  collision processes leading to the production of the ion. For the present charge states, only the initial (ground) state with $J_i=3/2$ of N-like ions can be aligned. 

The evaluation of the DC cross sections in Eq.~\eqref{eq:alignm_cross} was performed with the FAC code, which treats resonant electron capture in a distorted-wave framework \cite{gu2008}.  In this approach, the free electron is expanded in partial waves $(\varepsilon l _j)$ (e.g. \cite{csc1995,zhs2002, fss2009}) and the DC cross sections are traced back to the matrix elements $\left< \alpha_d J_d M_d \left|  V \right| \alpha_i J_i M_i(\varepsilon l_j)\right>$.   Here, the matrix element addresses the formation of a resonant state $\ket{ \alpha_d J_d}$ by resonant capture of a free electron by an initial ion with state $ \ket{\alpha_i J_i}$, through the electron-electron interaction $V$.  The influence of the Breit interaction can thus be investigated by performing calculations either with the full operator $V=V^{\mbox{\tiny C}}+V^{\mbox{\tiny B}}$, or with Coulomb interaction  only $V=V^{\mbox{\tiny C}}$. 

The allowed partial waves $(\varepsilon l_ j)$ are restricted by angular and parity selection rules included in the matrix elements. Moreover, since  $V$ is a two-body operator only matrix elements with two active pairs of orbitals participating in the DC process are nonzero \cite{joh2007}. 
%

 The DC process can be further reduced to a combination of Slater integrals between the radial components of two active pair-electrons. A Slater integral is given as 
\begin{eqnarray}
\Lambda_{x-v,y-w}^k&=&\int R_{x-v}(r_2)  dr_2 \int R_{y-w}(r_1)  \frac{r_{>}^k}{r_{<}^{k+1}}dr_1 \nonumber \\
&=&\int R_{x-v}(r_2) v^k(y,w,r_2)dr_2 ~,
%
\label{eq:radial_over} 
\end{eqnarray}
 where $R_{i-j}(r)$ can be either a density overlap, $R_{i-j}^{\tiny \mbox{C}}(r)=r^2\rho=P_i(r)P_jv(r)+Q_i(r)Q_j(r)$ resulted from  the Coulomb interaction, or can be overlaps resulting from the Breit interaction, which mixes the large ($P$) and small ($Q$) components of the radial wavefunctions, i.e., $R_{i-j}^{\tiny \mbox{B}}(r)=P_i(r)Q_j(r)\pm Q_i(r)P_j(r)$ \cite{grp1976, joh2007}.  
 The pairs $x-v$ and $y-w$ are the active orbitals that are changed during the DC process. 
 $r_{>}=\mbox{max}(r_2,r_1)$ and $r_{<}=\mbox{min}(r_2,r_1)$. Here, $m$ is a positive integer that depends on the angular decomposition of the matrix elements. 
As will be seen in Sec.~\ref{sec:Resul}, the observed state-selective influence of Breit interaction in DR angular distribution  can be traced back to the type of active orbitals and to the radial Coulomb and Breit overlaps between them.

\begin{figure*}[t]
  \centering
\includegraphics[clip=true,width=1.0\textwidth]{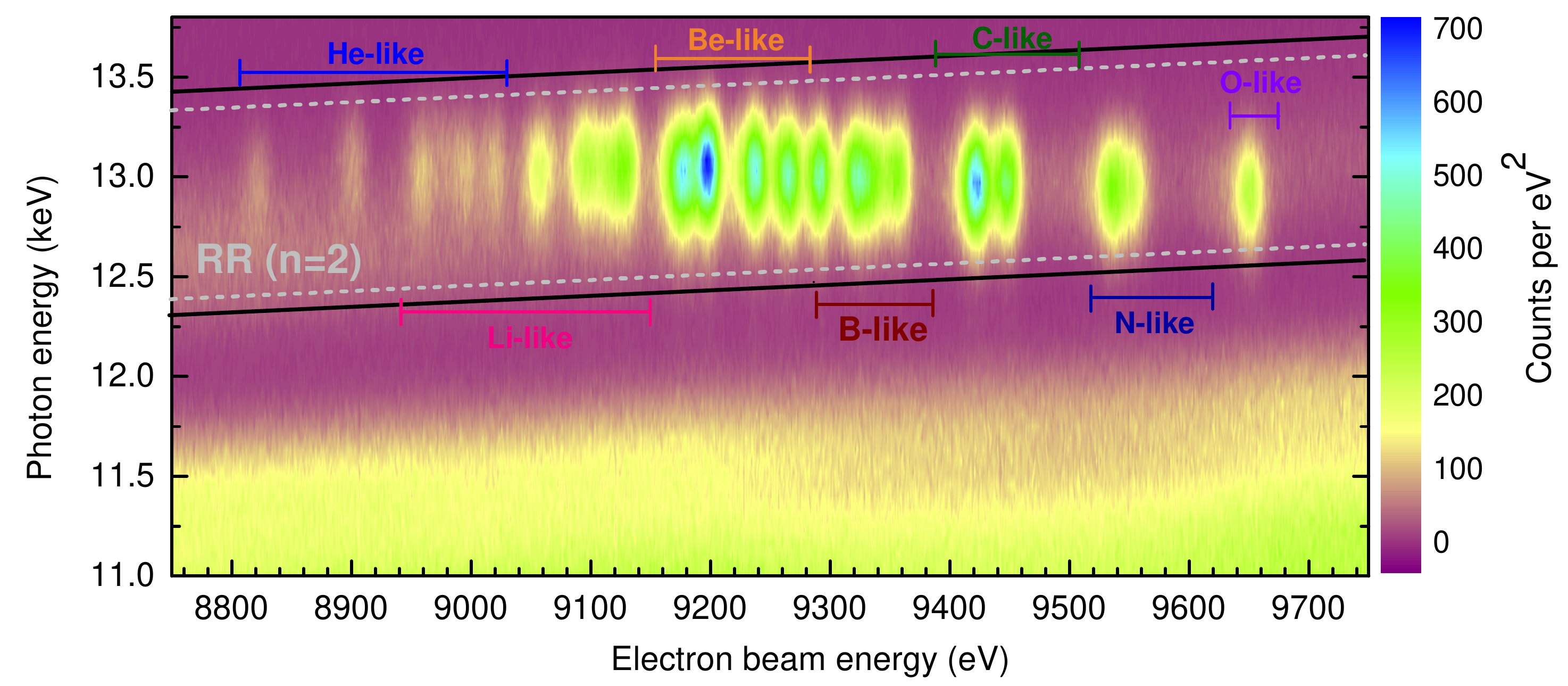}
\caption{ (Color online) Intensity of x rays in function  of x-ray energy and electron beam energy for the photon detector mounted at 90\degree to the electron beam direction. The solid lines delimitates the region of interest for the summation of events. The dashed lines delimitates the constant background of radiative recombination (RR) into $n=2$. Each bright spot is one DR resonance identified by the charge state of the initial ion. 
} 
\label{fig:exp_2D_spectr}
\end{figure*}
%

\section{Experiment}
\label{Experi_setup}
  

The experiment was performed at the Max-Planck Institut for Nuclear Physics, where highly charged ions of Kr were generated at the FLASH EBIT \cite{ecb2007,epp2010}.
In this experiment, an electron beam is emitted by an electron gun and accelerated towards a set of drift tubes with an applied high-voltage.
 This monoenergetic electron beam is simultaneously compressed by a magnetic field of \unit{6}{T} to a diameter of $\approx$\unit{50}{\mu m} (calculated according to Ref.~\cite{her1958}).
An electrostatic axial trap of \unit{50}{mm} length is formed by biasing the central drift tube with a slight positive voltage, relative to the two surrounding drift tubes. 
%
%
At the trap, injected atoms are then multi-ionized by the compressed electron beam through electron impact.
 A scheme of this experimental setup is shown in Fig.~\ref{fig:exp_ebit}.

 The electron beam energy was continuously scanned over the $KLL$ DR resonances by controlling the drift tube bias voltage according to a triangular wave function from \unit{6.7}{kV} to \unit{7.8}{kV} with a rate of \unit{1.8}{eV/s}. 
In order to limit the energy spread of the electron-ion interaction, the electron beam current was set to a moderate value of \unit{70}{mA}.
The cathode of the electron gun had a negative bias of -\unit{2}{kV}.

The trap settings were optimized for both  highest concentration of the highly charged ions and  energy beam resolution. The second criterion is accomplished by further reducing the energy spread with the  evaporative cooling technique \cite{pbl1991} through the use of a  shallow axial trap. However, shallow traps contains less ions than deeper ones, specially for He-like and Li-like ions. Therefore, we performed two types of measurements: (a) a deeper trap with higher concentration of high charge states (like He-like ions) and a  lower resolution of \unit{19}{eV} (full width at half maximum, FWHM); (b) A shallow trap with concentration of high charge states and a better resolution of \unit{12}{eV}. The respective values of the trap voltage offset for settings (a) and (b) are \unit{100}{V} and \unit{130}{V}.
%
%

High-purity germanium  detectors were mounted along and perpendicular to the electron beam axis,  as shown in Fig.~\ref{fig:exp_ebit}. 
The ratio between solid angles obtained from their location and active area is  $\Omega_{90^o}/\Omega_{0^o} \approx 9.4$.
%
%
 The photon energy resolution (FWMH) at \unit{13}{keV} of the detectors at 90\degree~and 0\degree~are \unit{180}{eV} and  \unit{200}{eV}, respectively.



\section{Data analysis}
\label{sec:data_analy}

A typical contour plot with the electron beam energy and photon energy as parameters is shown in Fig.~\ref{fig:exp_2D_spectr}. The electron beam energy was scanned through the  {\it KLL} DR resonances in He-like to O-like Kr, which are visible as bright spots. Each resonance is identified by the initial charge state of Kr before DR. The background near the resonances is due to radiative recombination (RR) into the $L$ shell ($n=2$), while the lower background (at photon energies of $\sim$11~keV) is due to radiative recombination in other heavier elements at the trap, like barium or tungsten. 

\begin{figure*}[t]
  \centering
\includegraphics[clip=true,width=\textwidth]{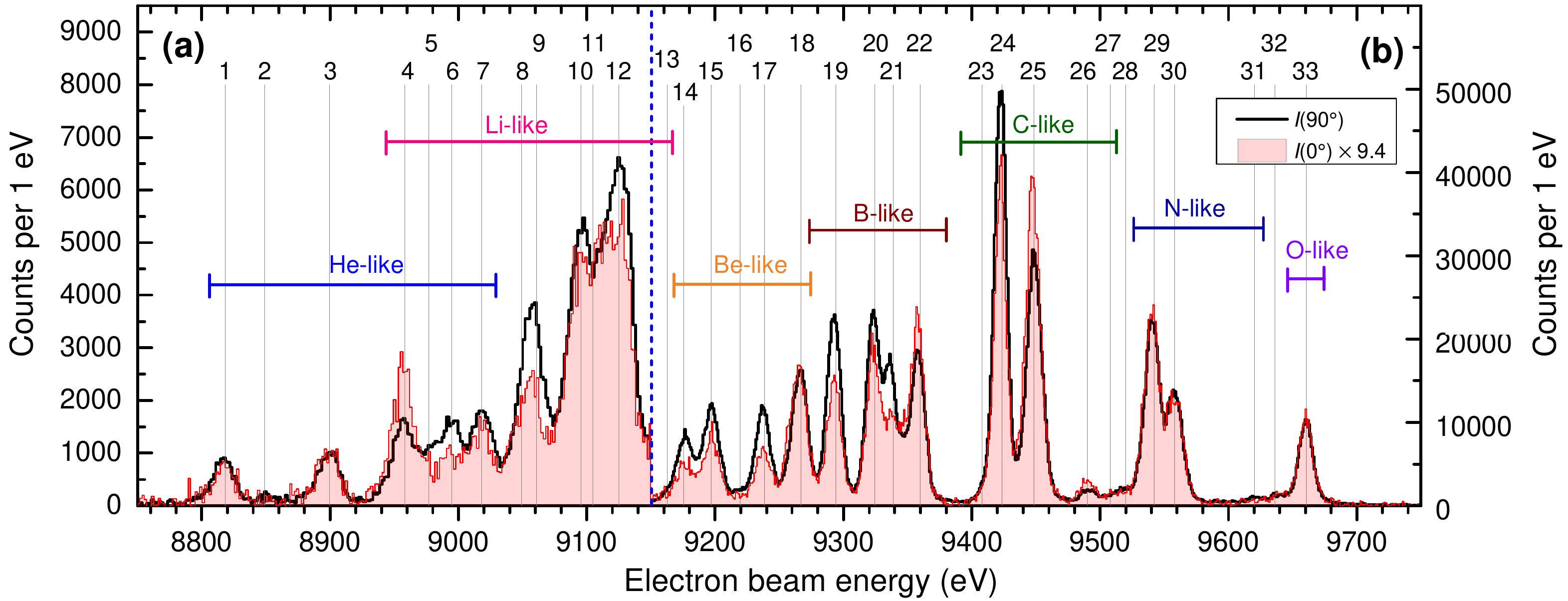}
\caption{(Color online) Intensity of DR x-rays. The black solid line corresponds to the observed perpendicular DR intensity. The red solid line shows the  DR intensity observed along the electron beam axis times  9.4. Each resonance is identified by one or more resonant states of a given charge state in Table~\ref{tab:res_ratio}. The left side (a) of the plot corresponds to a measurement performed with a deep trap  while the right side (b) corresponds to a shallow trap.
 } 
\label{fig:exp_DR_inten}
\end{figure*}

We obtained the intensity of DR x-rays by selecting the events of interest  of both RR and DR contained in the region of Fig.~\ref{fig:exp_2D_spectr} inside the black solid lines. These events were added up  for a given value of the electron beam energy. The obtained intensity of DR x-rays is displayed in Fig.~\ref{fig:exp_DR_inten}. Here, the background due to RR was fitted with a linear function and removed. The left (a) and right (b) sides of the plot correspond to a deep and shallow trap, respectively, as described in Sec.~\ref{Experi_setup}.

The selected energy region of Fig.~\ref{fig:exp_DR_inten} contains a high density of resonant states that were identified with the help of theoretical calculations.  Table~\ref{tab:res_ratio} displays the theoretical energies for the resonant states with respective identification, initial ion charge state and recombination process.
If two or more resonant states have the energy separated by less than half of the energy resolution,  we considered a single Gaussian function, for fitting the peak formed by these states. For now on, we refer to any of the obtained Gaussian functions as a resonance. The total number of  resonances is 33 as shown in Fig.~\ref{fig:exp_DR_inten}. The beam energy resolution of the settings (a) and (b) (see Fig.~\ref{fig:exp_DR_inten})  was obtained from resonances 3 and 24, each one consisted of a single well-separated resonant state. 
Some weak resonances, such as resonances 5, 8 or 11 have low resonant strengths, but contribute to the fit by perturbing the profiles of nearby, more intense resonances. 
The positions and widths of these resonances were fixed to theoretical energy values and to the previously obtained experimental width, having only the amplitudes as free-fit parameters. 
These resonances are identified in Table~\ref{tab:res_ratio} by not showing the experimental energy. 

 A linear calibration based on two theoretical resonances was employed for the electron beam energy. 
Measurement (a) in Fig.~\ref{fig:exp_DR_inten} was calibrated with resonances 3 and 24 (not shown), corresponding to energies of \unit{8899}{eV} and \unit{9427}{eV}, respectively. Similarly,  case (b) in Fig.~\ref{fig:exp_DR_inten} was calibrated with resonances 17 (\unit{9238}{eV}) and 24.  

    For each resonance we compare in Table~\ref{tab:res_ratio} the theoretical and experimental energy. The theoretical values  were obtained from the FAC code,  which follows a relativistic configuration interaction formalism (\cite{gu2008} and references therein). Other theoretical values based in MCDF and experimental values of resonant energies are available for DR resonances of C-like, N-like and O-like \cite{bpa2009} Kr and, more recently, for the case of H-like, He-like and Li-like ions \cite{hln2013}. We have a good agreement in all values of resonant energies with differences of less than \unit{4}{eV}.

Since the detectors at 0\degree~and 90\degree~have different solid angles of detection, we calibrated both spectra using an isotropic resonance. We choose resonance 33 that is the most intense resonance among the isotropic ones.  
The obtained solid angle correction factor due to the different solid angles was $C_{\Omega}=9.36\pm0.01$. 


  We quantified the emission anisotropy as the ratio between the DR resonance amplitudes observed along and  perpendicular directions to the electron beam axis, corrected with the solid angle factor, i.e.,  $\mathcal{R}=C_{\Omega} I(0\degree)/I(90\degree)$. Table~\ref{tab:res_ratio} list values of ratios $\mathcal{R}$ for each resonance along with the respective charge state of the initial ion and the recombination process that populated the resonant state.  
  %
%

\begin{figure}[t]
  \centering
\includegraphics[clip=true,width=1.0\columnwidth]{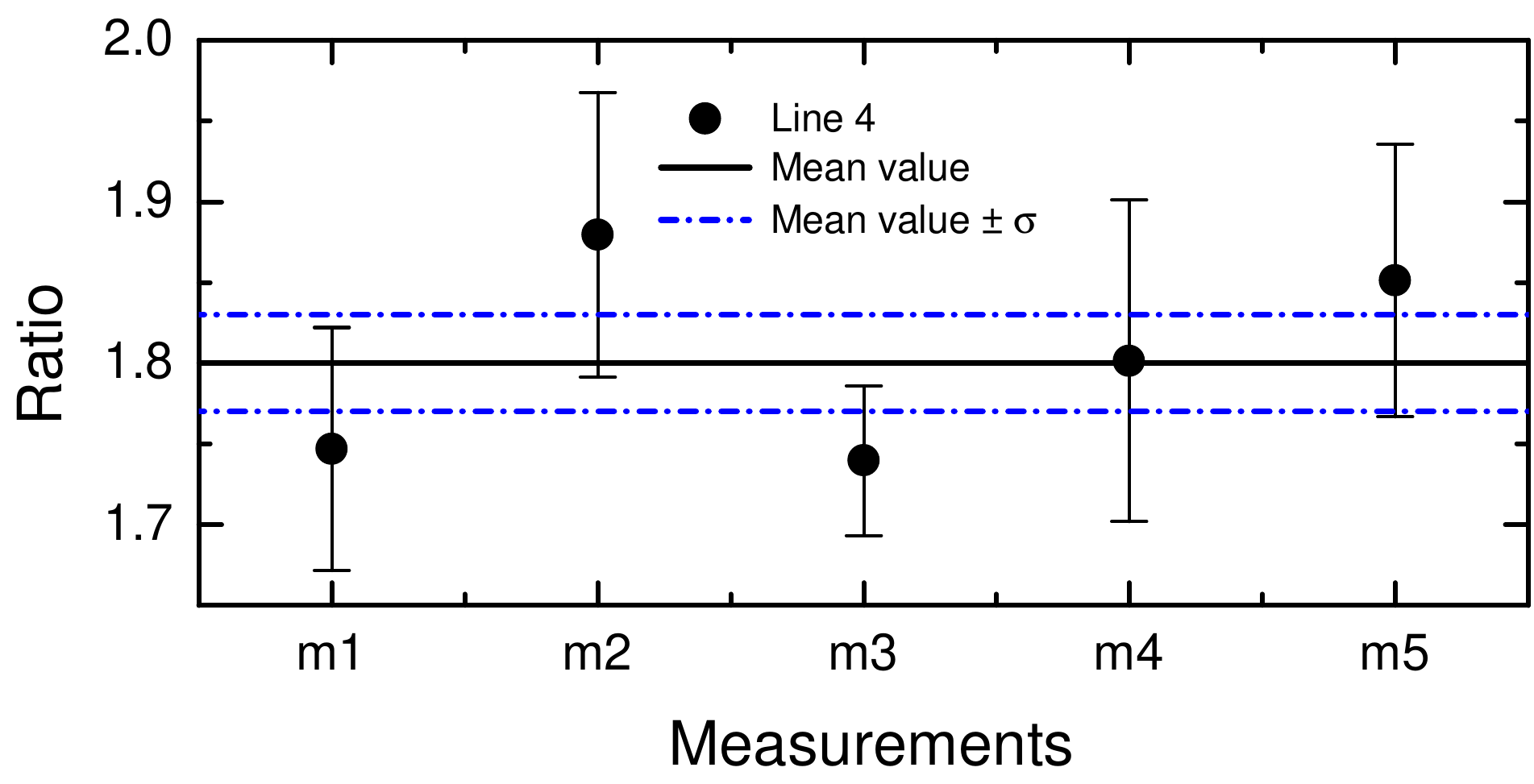} 
\caption{(Color online) Ratios of resonance 4 for several independent measurements. The black solid line corresponds to the weighted average and the blue dashed line stands for  plus or minus $\sigma$.
 } 
\label{fig:lines_measu}
\end{figure}

The obtained values of the ratios for the resonance 4 is shown in Fig.~\ref{fig:lines_measu}. The error bar in each measurement contains the combined uncertainty (1$\sigma$) of the statistical uncertainty of fitting the amplitudes for both $I(0\degree)$, $I(90\degree)$ and  $C_{\Omega}$.

According to the non-laminar optical model developed by Herrmann \cite{her1958}, the motion of an electron in an EBIT is described by an helical path collinear with the  beam direction. Thus, the relative electron-ion collision is not aligned  with the electron beam direction but  deviates by a pitch angle $\gamma$ that is given by \cite{bvr1996}
 \begin{equation}
 \tan \gamma = \frac{\sqrt{E_\bot }}{\sqrt{E_{\mbox{beam}} -E_\bot}}~,
 \label{eq:pitc_angl}
 \end{equation}
where $E_\bot$ is the transverse electron energy. 
%
 For a cathode temperature of \unit{1300}{K} and a cathode radius of \unit{1.5}{mm}, the transverse electron energy after compression of the electron beam is $E_\bot\approx$\unit{0.1}{keV}, which corresponds to a pitch angle of $\gamma \approx 6\degree$ \cite{bvr1996} and a deviation of 0.02 in $\mathcal{R}$. The final uncertainty of each $\mathcal{R}$ in Table~\ref{tab:res_ratio} is the combined uncertainty of the statistical error (see Fig.~\ref{fig:lines_measu}) and this systematic uncertainty.

\begin{table*}
\caption{\label{tab:res_ratio} Measured values of ratios $\mathcal{R}=C_{\Omega} I(0\degree)/I(90\degree)$ for $KLL$ resonances of highly charged Kr. The first column labels resonances in Fig.~\ref{fig:exp_DR_inten}, which can consist of a single resonant state or an ensemble of unresolved ones. The resonant states are given in $jj$-coupling notation, 
Column P identifies the resonant process and the charge state refers to the initial ion. 
Experimental and theoretical resonant energies are identified by ER$^{\mbox{exp}}$ and ER$^{\mbox{theo}}$, respectively. Theoretical values were obtained with FAC.
} 
\begin{ruledtabular} 
\begin{tabular}{llllllr}
Label&  P&Charge state & Resonant state   & ER$^{\mbox{exp}}$&ER$^{\mbox{theo}}$&  $\mathcal{R}\phantom{00}$  \\
\\[-2.0ex] \hline  \\[-2.0ex] 
1		&	DR	&		He		&		$[(1s(2s^2)_0]_{1/2}$			&		8821.5 		$\pm$0.4	&		8820		&	0.99	$\pm$	0.07	\\
2		&	DR	&		He		&		$[(1s2s)_12p_{1/2}]_{3/2}$			&		8849\phantom{.0}	$\pm$4	&		8849		&	0.2\phantom{0}		$\pm$	0.3\phantom{0}	\\
3		&		DR		&		He		&		$[(1s2s)_02p_{1/2}]_{1/2}$			&		calib.			&		8899  		&			1.01 	$\pm$	0.06	\\
4		&		DR		&		Li		&		$[1s2s^22p_{1/2}]_1$			&		8953	\phantom{.} 	$\pm$1	&		8954 		&	1.80	$\pm$	0.04		\\
5		&		DR		&		He		&		$[(1s2s)_12p_{3/2}]_{1/2}$			&	blend		&		8968		&					\\
		&		DR		&		He		&		$[(1s2s)_02p_{3/2}]_{3/2}$			&		 		&		8977		&													\\
6		&		DR		&		He		&		$[(1s2p_{1/2})_12p_{3/2}]_{5/2}$			&		8990	\phantom{.} 	$\pm$1	&		8989		&	0.59	$\pm$	0.09	 \\
7		&		DR		&		He		&		$[(1s2p_{1/2})_02p_{3/2}]_{3/2}$			&		9014.8	$\pm$0.5	&		9013 		&			0.82		$\pm$	0.07	\\
		&		DR		&		Li		&		$[1s2s^22p_{1/2}]_2$			&		 			&		9014 		&													\\
8		&		DR		&		Li		&		$[((1s2s)_12p_{1/2})_{3/2}2p_{3/2}]_3$			&	blend		&		9048 		&				\\
9		&		DR		&		He		&		$[1s(2p_{3/2}^2)_2]_{5/2}$			&	blend		&		9058			&				\\
10		&		DR		&		Li		&		$[((1s2s)_12p_{1/2})_{3/2}2p_{3/2}]_2$			&		9092.0		$\pm$0.4	&		9092 		&	0.80	$\pm$	0.04	\\
11		&		DR		&		Li		&		$[((1s2s)_12p_{1/2})_{3/2}2p_{3/2}]_1$			&	blend		&		9105 		&				\\
12		&		DR		&		Li		&		$[((1s2s)_02p_{1/2})_{1/2}2p_{3/2}]_2$			&		9126.0	$\pm$0.3	&		9121 		&			0.81	$\pm$	0.03	\\
		&		DR		&		Li		&		$[(1s2s)_1(2p_{3/2})_2^2]_3$			&		 			&		9125		&													\\
13		&		DR		&		Li		&		$[(1s2s)_1(2p_{3/2})_2^2]_{2}$			&	blend		&		9164		&				\\
14		&		DR		&		Be		&		$[(1s2s^22p_{1/2})_12p_{3/2}]_{5/2}$			&		9174.9	$\pm$0.2	&		9171 		&	0.57	$\pm$	0.03	\\
15		&		DR		&		Be		&		$[(1s2s^22p_{1/2})_12p_{3/2}]_{3/2}$			&		9196.4	$\pm$0.2	&		9194		&	0.62	$\pm$	0.02	\\
16		&		DR		&		Li		&		$[(1s2s)_0(2p_{3/2})_0^2]_0$			&	blend		&		9220 		&				\\
17		&		DR		&		Be		&		$[1s2s^2(2p_{3/2}^2)_2]_{5/2}$			&		calib.			&		9238 		&	0.55	$\pm$	0.02	\\
18		&		DR		&		Be		&		$[1s2s^2(2p_{3/2}^2)_2]_{3/2}$			&		9267.5	$\pm$0.2	&		9260 		&	1.04		$\pm$	0.02	\\
		&		DR		&		B		&		$[1s2s^2(2p_{1/2})^22p_{3/2}]_2$			&					&		9265 		&													\\
		&		DR		&		Be		&		$[1s2s^2(2p_{3/2}^2)_0]_{1/2}$			&					&		9271 		&													\\
19		&		DR		&		B		&		$[1s2s^2(2p_{1/2})^22p_{3/2}]_1$			&		9295.5	$\pm$0.2	&		9293		&	0.61	$\pm$	0.02	\\
20		&		DR		&		B		&		$[(1s2s^22p_{1/2})_1(2p_{3/2}^2)_2]_2$			&		9326.0	$\pm$0.3	&		9324		&	0.79	$\pm$	0.02	\\
21		&		DR		&		B		&		$[(1s2s^22p_{1/2})_1(2p_{3/2}^2)_2]_3$			&		9344.1	$\pm$0.2	&		9337 		&	0.82	$\pm$	0.03	\\
		&		DR		&		B		&		$[(1s2s^22p_{1/2})_1(2p_{3/2}^2)_2]_1$			&					&		9345 		&													\\
22		&		DR		&		B		&		$[(1s2s^22p_{1/2})_0(2p_{3/2}^2)_2]_2$			&		9362.1	$\pm$0.2	&		9359 		&	1.15	$\pm$	0.02	\\
		&		DR		&		B		&		$[(1s2s^22p_{1/2})_1(2p_{3/2}^2)_0]_1$			&					&		9362		&			    										\\
23		&		TR		&		B		&		$[1s2s^22p_{3/2}^3]_{2}$			&	blend		&		9408		&				\\
24		&		DR		&		C		&		$[1s2s^22p_{1/2}^2(2p_{3/2}^2)_2]_{5/2}$			&		calib.			&		9427		&	0.73	$\pm$	0.02	\\
25		&		DR		&		C		&		$[1s2s^22p_{1/2}^2(2p_{3/2}^2)_2]_{3/2}$			&		9453.0	$\pm$0.1	&		9452 		&	1.21	$\pm$	0.02	\\
		&		DR		&		C		&		$[1s2s^22p_{1/2}^2(2p_{3/2}^2)_0]_{1/2}$			&		 			&		9454 		&													\\
26		&		TR		&		C		&		$[(1s2s^22p_{1/2})_02p_{3/2}^3]_{3/2}$			&		9493.6 		$\pm$0.8	&		9494 		&				\\
		&		QR		&		Be		&		$[(1s2p_{1/2})_02p_{3/2}^3]_{3/2}$			&					&		9495 		&		 											\\					
27		&		TR		&		C		&		$[(1s2s^22p_{1/2})_12p_{3/2}^3]_{5/2}$			&	blend		&		9503		&				\\
		&		TR		&		C		&		$[(1s2s^22p_{1/2})_12p_{3/2}^3]_{1/2}$			&		 			&		9507		&													\\			
28		&		QR		&		Be		&		$[(1s2p_{1/2})_12p_{3/2}^3]_{1/2}$			&	blend		&		9516 		&								\\
		&		QR		&		Be		&		$[(1s2p_{1/2})_12p_{3/2}^3]_{3/2}$			&		 		&		9524 		&													\\
29		&		DR		&		N		&		$[1s2s^2(2p_{1/2})^22p_{3/2}^3]_2$			&		9540.7 $\pm$0.2	&		9541		&				1.02		$\pm$	0.02	\\
30		&		DR		&		N		&		$[1s2s^2(2p_{1/2})^22p_{3/2}^3]_1$			&		9557.2 	$\pm$0.2	&		9558 		&				1.00  		$\pm$	0.02	\\
31		&		TR		&		C		&		$[(1s2s)_1(2p_{1/2})^2p_{3/2}^3]_{3/2}$			&	blend		&		9626 		&								\\
		&		QR		&		B		&		$[(1s2s)_0 p_{3/2}^4]_{0}$			&		 			&		9633		&													\\
32		&		QR		&		B		&		$[1s2p_{1/2}^22p_{3/2}^3]_{2}$			&	blend		&		9656		&													\\
33		&		DR		&		O		&		$[1s2s^22p_{1/2}^2 2p_{3/2}^4]_{1/2}$			&					&		9653 		&													\\
\end{tabular}
\end{ruledtabular}
\end{table*}

   
   %
%

 %
 
   Besides the (usual) DR process, we also observe resonant states populated by higher-order resonant recombination, such as trielectronic (TR) and quadruelectronic (QR) recombination processes \cite{bpa2009, bmb2011} as identified by some resonances in Table~\ref{tab:res_ratio}.  In the present work we restrict to their energy identification. Investigations of influence of angular distribution and polarization properties to hot plasma model have been published elsewhere \cite{sas2016}.
   %





 %
 %


\begin{table*}
\caption{\label{tab:res_alignm}  Values the experimental alignment parameter $\mathcal{A}_2^{\mbox{exp}}$  for a selection of resonant states. Intermediate states are identified by a label  and  total angular momentum $J_d$. Allowed partial waves of the free electron are displayed in the third column. Total radiative decay widths $W_{r}^{df}$ calculated with FAC are listed in the fifth column. The term $\mathcal{T}$ in the sixth column denotes the average value of $\alpha_{df}$ weighted by $W_{r}^{df}$. Theoretical values of $\mathcal{A}_2^{\mbox{theo}}$ are displayed for the cases of including BI and only Coulomb interaction (C).
} 
\begin{ruledtabular} 
\begin{tabular}{llllllllllllll}
Label  & $J_d$ &Partial wave & Final state& $W_r^{df}$ (s$^{-1}$)&$\mathcal{T}$&$\mathcal{A}_2^{\mbox{exp}}$ & $\mathcal{A}_2^{\mbox{theo}}$ (C) & $\mathcal{A}_2^{\mbox{theo}}$ (C+BI) \\  
\\[-2.0ex] \hline  \\[-2.0ex]
4	&	1	&		$\varepsilon p_{1/2},~ \varepsilon p_{3/2}$	&	$[1s^22s^2]_0$	&	3.591(14)	&	0.682	&	0.62	$\pm$0.02	&	0.704	&	0.640		\\
	&		&			&	$[1s^2(2p^2_{1/2})_0]_0$	&	1.235(13)	&		&			&		&			\\
	&		&			&	$[1s^2(2p_{1/2}2p_{3/2})_1]_1$	&	3.452(12)	&		&			&		&			\\
	&		&			&	$[1s^2(2p_{1/2}2p_{3/2})_2]_2$	&	8.865(12)	&		&			&		&			\\
	\\[-2.0ex]  \cline{4-9} \\[-2.0ex]																			
19	&	1	&		$\varepsilon s_{1/2},~ \varepsilon d_{3/2}$	&	$[1s^22s^22p^2_{1/2}]_0$	&	1.138(15)	&	0.453	&	-0.66$\pm$0.04	&	-0.694	&	-0.681		\\
	&		&			&	$[1s^2 2s^2 (2p_{1/2}2p_{3/2})_1]_1$	&	1.142(14)	&		&			&		&			\\
	&		&			&	$[1s^2 2s^2 (2p_{1/2}2p_{3/2})_2]_2$	&	5.145(14)	&		&			&		&			\\
	&		&			&	$[1s^2 2s^2(2p^2_{3/2})_2]_2$	&	2.771(12)	&		&			&		&			\\
	\\[-2.0ex]   \hline \\[-2.0ex]																			
15	&	3/2	&		$ \varepsilon d_{3/2}$	&	$[1s^2 2s^2 2p_{1/2}]_{1/2}$	&	1.163(15)	&	0.318	&	-0.91$\pm$0.05	&	-1.000	&	-1.000		\\
	&		&			&	$[1s^2 2s^2 2p_{1/2}]_{3/2}$	&	2.861(14)	&		&			&		&			\\
	&		&			&	$[1s^2  2p_{3/2}^3]_{3/2}$	&	3.599(12)	&		&			&		&			\\
	&		&			&	$[1s^2 2p_{1/2} (2p_{3/2}^2)_2]_{3/2}$	&	2.978(12)	&		&			&		&			\\
	&		&			&	$[1s^2 2p_{1/2} (2p_{3/2}^2)_2]_{5/2}$	&	8.662(12)	&		&			&		&			\\
	\\[-2.0ex]   \hline \\[-2.0ex]																			
10	&	2	&		$\varepsilon d_{3/2},~\varepsilon d_{5/2}$	&	$[1s^2(2s 2p_{1/2})_1]_1$	&	1.192(15)	&	0.270	&	-0.5\phantom{0}$\pm$0.1	&	-0.604	&	-0.604		\\
	&		&			&	$[1s^2(2s 2p_{3/2})_2]_2$	&	1.646(14)	&		&			&		&			\\
	&		&			&	$[1s^2(2s 2p_{3/2})_1]_1$	&	9.222(13)	&		&			&		&			\\
	\\[-2.0ex]  \cline{4-9} \\[-2.0ex]																			
20	&	2	&		$\varepsilon d_{3/2},~\varepsilon d_{5/2}$	&	$[1s^2 2s^2 (2p_{1/2}2p_{3/2})_1]_1$	&	5.114(14)	&	0.179	&	-0.84	$\pm$0.09	&	-0.929	&	-0.939		\\
	&		&			&	$[1s^2 2s^2 (2p_{1/2}2p_{3/2})_2]_2$	&	7.714(13)	&		&			&		&			\\
	&		&			&	$[1s^2 2s^2(2p^2_{3/2})_2]_2$	&	1.272(14)	&		&			&		&			\\
	\\[-2.0ex]   \hline \\[-2.0ex]																			
6	&	5/2	&		$\varepsilon d_{5/2}$	&	$[1s^2 2p_{3/2}]_{3/2}$	&	4.076(14)	&	0.374	&	-0.9\phantom{0}$\pm$0.2	&	-1.069	&	-1.069		\\
	\\[-2.0ex]  \cline{4-9} \\[-2.0ex]																			
14	&	5/2	&		$\varepsilon d_{5/2}$	&	$[1s^2  2p_{3/2}^3]_{3/2}$	&	3.737(14)	&	0.354	&	-0.95$\pm$0.08	&	-1.069	&	-1.069		\\
	&		&			&	$[1s^2 2p_{1/2} (2p_{3/2}^2)_2]_{5/2}$	&	3.012(12)	&		&			&		&			\\
	\\[-2.0ex]  \cline{4-9} \\[-2.0ex]																			
17	&	5/2	&		$\varepsilon d_{5/2}$	&	$[1s^2  2p_{3/2}^3]_{3/2}$	&	5.128(14)	&	0.357	&	-0.99$\pm$0.05	&	-1.069	&	-1.069		\\
	&		&			&	$[1s^2 2p_{1/2} (2p_{3/2}^2)_2]_{5/2}$	&	1.122(13)	&		&			&		&			\\
	\\[-2.0ex]  \cline{4-9} \\[-2.0ex]																			
24	&	5/2	&		$\varepsilon d_{5/2}$	&	$[1s^2 2s^2 2p^2_{1/2}2p_{3/2}]_{3/2}$	&	4.667(14)	&	0.210	&	-0.94	$\pm$0.08	&	-1.069	&	-1.069		\\
	&		&			&	$[1s^2 2s^2 2p_{1/2}(2p^2_{3/2})_2]_{3/2}$	&	3.190(14)	&		&			&		&			\\
	&		&			&	$[1s^2 2s^2 2p_{1/2}(2p^2_{3/2})_2]_{5/2}$	&	2.023(14)	&		&			&		&			\\
  \end{tabular}
\end{ruledtabular}
\end{table*}

\section{Results and discussion}
\label{sec:Resul}



The obtained values $\mathcal{R}=$0.99$\pm0.07$ and $\mathcal{R}=$1.01$\pm0.06$  for the isotropic resonances 1 and 3 listed in Table~\ref{tab:res_ratio} are in agreement with the ones expected from an  isotropic emission of these resonances ($J_d=1/2$).

From all resonances listed in Table~\ref{tab:res_ratio}, we observe that the majority of them have $\mathcal{R}<1$, i.e, the photons are mostly emitted in a direction perpendicular  to the electron beam axis. According to Eq.~\eqref{eq:ang_dis} this is equivalent to the product $\alpha_{d\rightarrow f}^2 \mathcal{A}_2^d$ being negative. Due to radiative and collision deexitation, the ions are mostly populated in the ground state, which for He-like, Be-like, and C-like ions have $J_i=0$, while for Li-like and B-like ions have $J_i=1/2$. Therefore, for He-like, Be-like, and C-like ions, the intermediate magnetic sublevel states  are limited to $M_d=\pm1/2$. This leads to an orientation of $J_d$ being mostly perpendicular to the electron trajectory, and thus to a negative value of $\mathcal{A}_2^d$.
For  Li-like and B-like ions,  a stronger photon emission being perpendicular to the electron beam axis and $\alpha_{d\rightarrow f}^2>0$  indicates that  magnetic sublevels $M_d=0$ are more populated.  
 Some observed resonances have $\mathcal{R}>1$. This is due to $\alpha_{df}<0$ for photon transitions with $J_d=J_f$.  Such resonances include 22, 25, 26, 29 and 30. Therefore, the rule-of-thumb is that resonant magnetic sublevels with $M_d =0$ and $M_d=\pm1/2$ are mostly populated. 
 This can be expected, since the orbital angular momentum of the free electron, which is perpendicular to its trajectory, is  transferred in the collision. This results in an orientation of $J_d$ that is mostly perpendicular to the electron beam axis. 

Resonance 4 is an exception to this general observation by having $\mathcal{R}>1$ and $\alpha_{df}>0$, indicating that magnetic sublevels with $M_d=\pm1$ are predominantly populated.  This is a consequence of the sublevel $M_d=0$ being weakly populated due to LS selection rules of the term  $^3P_1$ \cite{fss2009}.

For the investigation of the BI influence we restricted ourselves to a set of resonances with one well-separated  resonant state. 
Moreover, resonances of initial N-like ions were also not considered here due to a possible alignment  of the initial state (Sec.~\ref{sec:theor}).
%
The list of selected resonant states sorted by $J_d$ is given in Table~\ref{tab:res_alignm} together with experimental and theoretical values of $\mathcal{A}_2^{d}$. The experimental values of $\mathcal{A}_2^{d}$ were extracted from the ratios $\mathcal{R}$, listed in Table~\ref{tab:res_ratio}, according to the following expression
 \begin{equation}
\mathcal{A}_2^d = \frac{\mathcal{R}-1 }{\mathcal{T}\left(\frac{\mathcal{R}}{2}+1\right)}   ~, 
%
\label{eq:ratio_df} 
\end{equation}
which can be obtained from Eq.~\eqref{eq:ang_dis}. Here, the term $\mathcal{T}$ stands for the average of $\alpha_{d\rightarrow f}$ weighted by the radiative decay rates, i.e.,  $\mathcal{T}= \sum_{f}  W_r^{df} \alpha_{d\rightarrow f} /\sum_{f}  W_r^{df}. $

%

Theoretical values of $\mathcal{A}_2^{d}$ were calculated according to Eq.~\eqref{eq:alignm_cross} with the DC cross sections obtained from FAC, with and without inclusion of Breit interaction in the calculations.  Table~\ref{tab:res_alignm} also lists the allowed partial waves that describe the free electron for the DR process of each resonant state, alongside the radiative decay rate $W_r^{df}$ (calculated with FAC) to the final states necessary for calculating $\mathcal{T}$.   By comparing the results of $\mathcal{A}_2 ^d $, calculated with the DC cross sections  of FAC, with all respective values in Ref.~\cite{fss2009}, we noticed maximum differences of 5~\% along all the isoelectronic sequence. 
 Experimental results presented in Table~\ref{tab:res_alignm} agree with the theoretical predictions within 1.5$\sigma$ for all resonances. 

 %
  \begin{figure}[t]
  \centering
\includegraphics[clip=true,width=1.0\columnwidth]{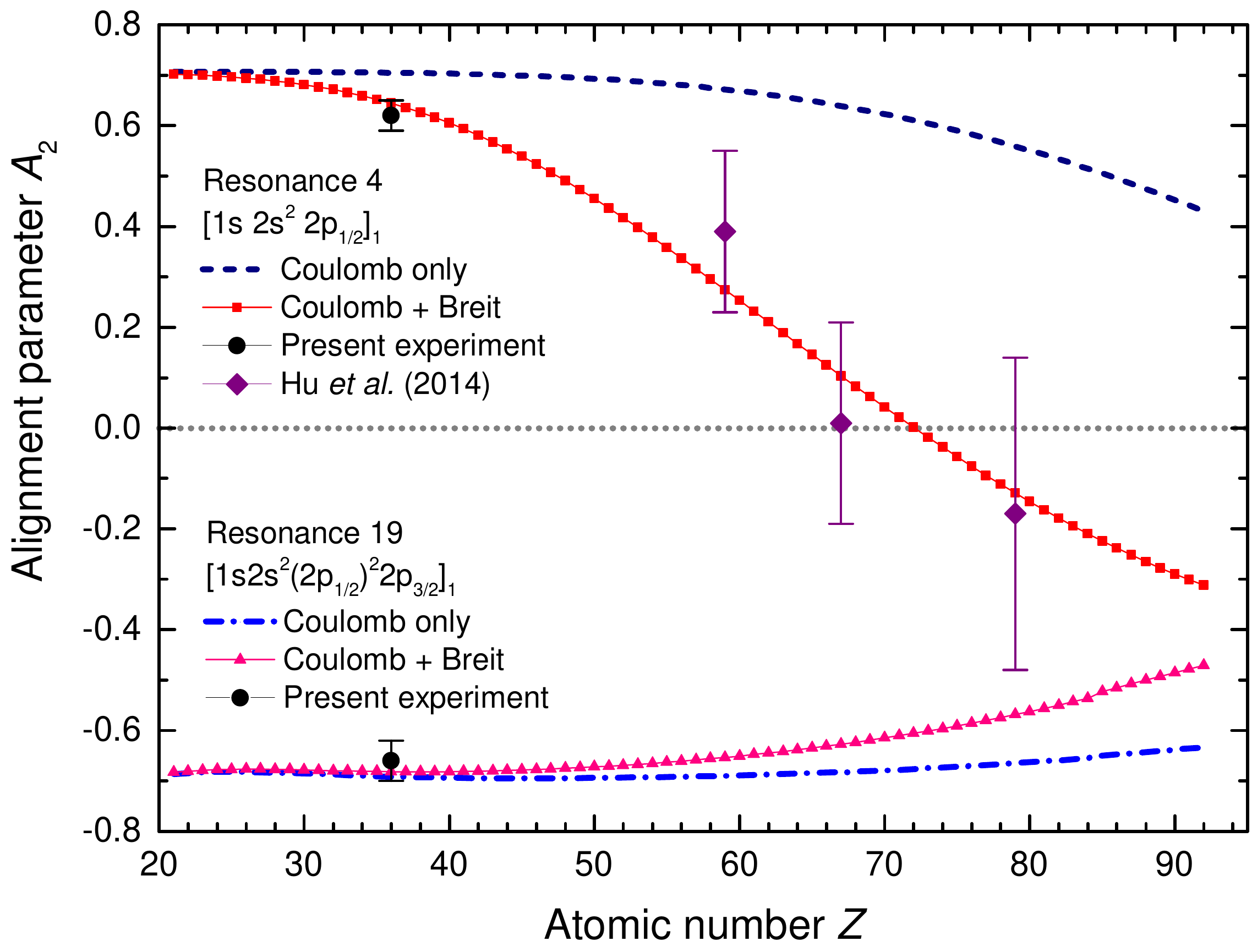}   
\caption{(Color online)  Alignment parameter $\mathcal{A}_2^{d}$ for the intermediate states $[1s2s^22p_{1/2}]_1$  (resonance 4) and $[1s2s^2(2p_{1/2})^22p_{3/2}]_1$ (resonance 19).  Theoretical values were evaluated with (C+BI) and without Breit  interaction (C), according to Eq.~\eqref{eq:alignm_cross}. The present experimental results are represented by the dark circles, while values provided by Refs.~\cite{hhl2012, hlh2014} for heavier elements are represented by the purple diamonds.
 } 
\label{fig:Breit_infle}
\end{figure}

      \begin{figure*}[t]
  \centering
\includegraphics[clip=true,width=\textwidth]{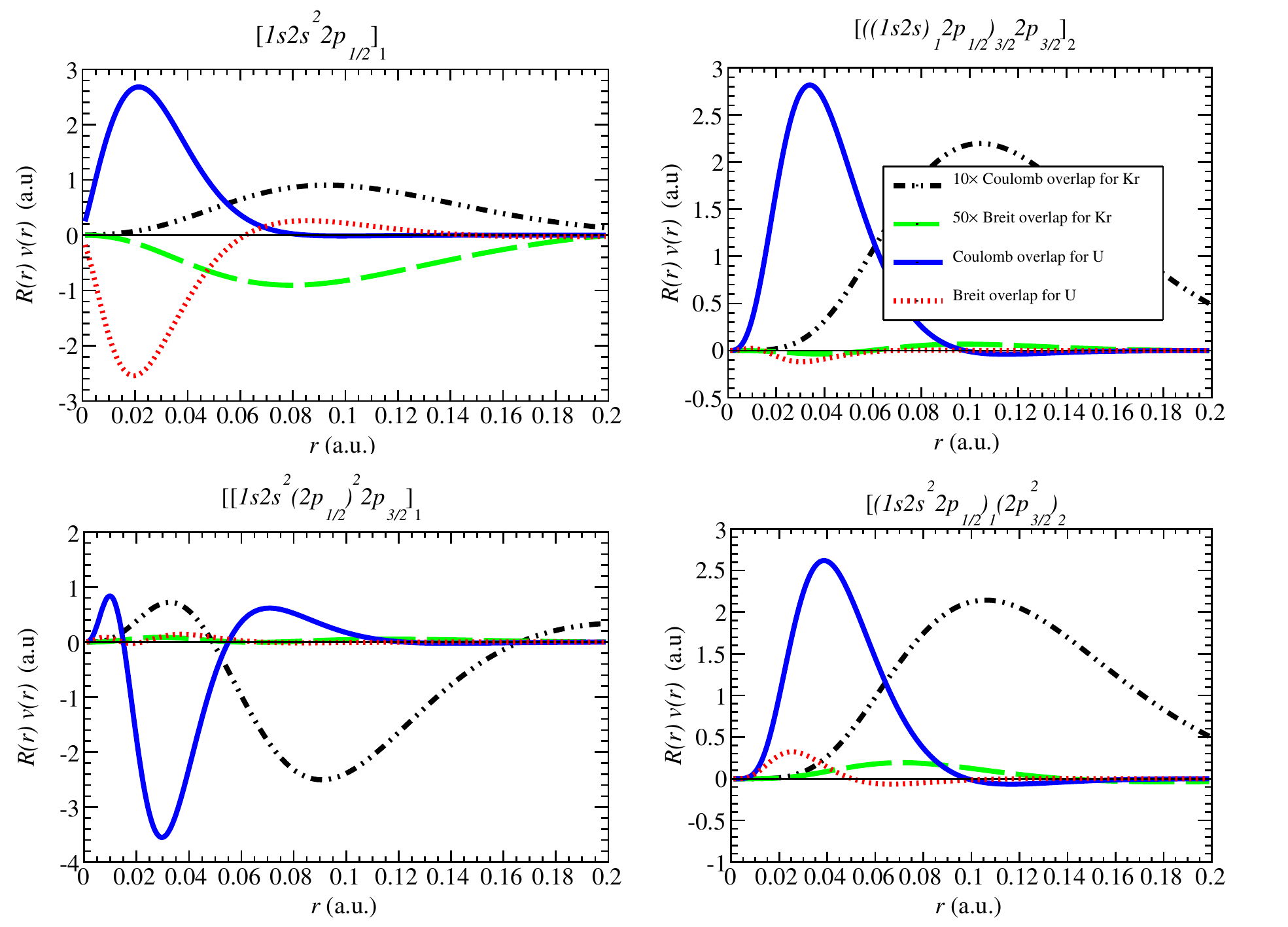} 
\caption{(Color online) Coulomb and Breit overlaps of active orbitals for resonances 4, 10, 19 and 20, as well as for Kr and U ions. The  overlaps \eqref{eq:radial_over} are $R_{2p_{1/2}-\varepsilon p_{1/2}}(r) v(1s, 2s, r)$, $R_{2p_{3/2}-\varepsilon d_{3/2}}(r) v(1s, 2p_{1/2}, r)$, $R_{2p_{3/2}-\varepsilon s}(r) v(1s, 2p_{1/2}, r)$ and $R_{2p_{3/2}-\varepsilon d_{3/2}}(r) v(1s, 2p_{3/2}, r)$ for resonances 4, 10, 19 and 20, respectively.  Atomic units are used.
%
 } 
\label{fig:acti_electr}
\end{figure*}

Initial He-like, Be-like and C-like ions with $J_i=0$ have the DC process restricted to one partial wave, due to selection rules, which reduces the alignment parameter to a geometrical factor \cite{csc1995}. Exact values of $\mathcal{A}_2 ^d $  are $-1$ and $-4\sqrt{1/14}\approx -1.069$ for $J_d=3/2$ and $J_d=5/2$, respectively. 
On the other hand,  Li-like and B-like ions include two partial waves, and the interference between them makes the $\mathcal{A}_2^{d}$ dependent of the details of the electron-electron interaction \cite{csc1995,fsg2012}. As can be observed in Table~\ref{tab:res_alignm}, $\mathcal{A}_2^{d}$ of resonant states 4, 19 and 20 depends on the  contribution of BI. The resonant state 10 also contains a dependency that is smaller than the uncertainty  shown in Table~\ref{tab:res_alignm}.  
%
We observed that without the contribution of Breit interaction, the difference between experimental and theoretical values amounts 3.5$\sigma$.  
On the other hand, BI influence is too small to be observed in resonances 10, 19 and 20, which agree with predictions. 
As depicted in Fig,~\ref{fig:Breit_infle}, even for an heavy element like U, the influence of  BI can be regarded as a correction for the resonance 19.   A pronounced influence of BI was predict for the resonant state 4 \cite{fss2009} and experimentally demonstrated for  heavy elements \cite{hhl2012, hlh2014}. 
%
%
 %
%
%

For investigation of the state-selective influence of BI, we calculated the radial overlaps in Eq.~\eqref{eq:radial_over} between the active electrons participating in the DC process. 
  The respective radial overlaps were calculated with FAC  for the resonant states 4, 10, 19 and 20, as well as for Kr and U.  The result is displayed in Fig.~\ref{fig:acti_electr} for the cases of  Breit and Coulomb radial overlaps. The integer $k$ of Eq.~\eqref{eq:radial_over} for these overlaps is the minimal value allowed by selection rules. For Coulomb overlaps it is $k=1$ in all cases, with the except of the case 4 (with $k=0$), whereas for Breit overlap it is $k=0$ for all cases. 
%
%
     
  It can be noticed for the resonance 4 that  the Breit overlap is similar in magnitude to the Coulomb overlap by factors of 5 and 1, for Kr and U ions, respectively. On the other hand, for all other resonances the Breit overlap is regarded as a small correction to the Coulomb overlap.
  In the case of resonance 4, all active orbitals $1s$, $2s$ and $2p_{1/2}$ have their relativistic small components and large components mostly located in the same radial region ($\approx a_0/Z$ with $a_0$ being the Bohr radius).
This is not the case for the other resonant states, where the small and large components of the active orbitals $1s$, $2p_{1/2}$ and $2p_{3/2}$  are located in different radial regions, the last one mostly centered at $\approx 4 a_0/Z$. This delocalization of the radial large and small components reduces the Breit overlap and makes it much smaller than the Coulomb overlap.  
   
Therefore, the relative importance of the Breit interaction in the alignment of doubly excited states is mostly related to the radial localizations of the large and small components of the relativistic wavefunctions.

%


%
\section{Summary}
\label{sec:sum}



In this work we performed a systematic investigation of the angular distribution of the emitted photons produced by $KLL$ DR of highly charged ion, from He-like to O-like Kr.  The  radiation was recorded along and perpendicular to the electron beam axis. Experimental alignment parameters were extracted from the data.  
%
%
 Among the extensive set of observed resonant states, only one  manifests an observable dependency of Breit interaction on the alignment parameter. The lack of dependency of BI by the other resonances can be traced back to the radial overlaps between the active electrons of the DC process. 

%
\begin{acknowledgments}

This work was supported by the German Research Foundation (DFG) within the Emmy Noether program under Contract No. TA 740 1-1. 
%

\end{acknowledgments}

\bibliography{DR_articles}


\end{document}